\def\fun#1#2{\lower3.6pt\vbox{\baselineskip0pt\lineskip.9pt
  \ialign{$\mathsurround=0pt#1\hfil##\hfil$\crcr#2\crcr\sim\crcr}}}
\relax
\documentstyle[12pt]{article}
\advance\textheight by \topskip
\begin{document}
\rightline{\bf ILG-TMP-93-01}
\rightline{\bf February, 1993}
\bigskip
\bigskip
\bigskip
\centerline{{\bf \large Alexander A.Belov}\footnote{ e-mail:
mitpan@adonis.ias.msk.su [Belov]}}
\smallskip
\centerline{Int.Inst.for Math.Geophys.and Predictions Theory}
\centerline{Warshavskoe sh. 79, k.2 Moscow 113556 Russia}
\bigskip
\bigskip
\centerline{{\bf \large Karen
D.Chaltikian}\footnote{ e-mail: mitpan@adonis.ias.msk.su,
itf@ips.ac.msk.su [Chaltikian]}}
\smallskip
\centerline{Landau Institute for Theoretical Physics}
\centerline{Vorobyevskoe sh.2, Moscow 117940 Russia}
\bigskip
\bigskip
\centerline{\bf \large Lattice analogues of W-algebras and}
\medskip
\centerline{\bf \large Classical Integrable Equations}
\bigskip
\bigskip
\bigskip
\centerline{\bf Abstract}
\bigskip

We propose a regular way to construct lattice versions of $W$-algebras,
both for quantum and classical cases. In the classical case we write the
algebra explicitly and derive the lattice analogue of Boussinesq equation
from the Hamiltonian equations of motion. Connection between the lattice
Faddeev-Takhtadjan-Volkov algebra [1] and q-deformed Virasoro is also
discussed.

\bigskip
\bigskip
\bigskip

\def\qbinom#1#2{\left[\matrix{#1\cr #2\cr}\right]}
\section{ Introduction}
\bigskip

     Recently there has been a  great  interest  in  the  lattice
analogues of 2D field  theoretical  models  possessing  conformal
invariance [1-4]. For the first  time  this  interest  arised  in
connection with Liouville model [1-3].  As  is  well  known,  its
quantization in continuous limit is rather difficult in the range
of central charges $1<c<25$, where  usual  perturbation  theory  is
inapplicable. 'Latticization' of the world sheet can be viewed as
some   alternative   to   usual   point-splitting    method    of
regularization.  That's  why  lattice  conformal  theories   give
another way of quantization, more convenient  in  some  cases  in
comparison with the that of based on normal  ordering  procedure.
An additional motivation for studying  the  lattice  analogue  of
conformal invariance comes from the traditional understanding  of
Integrable  Massive  models  as   appropriate   deformations   of
Conformal  ones  [5-7].  It  seems  highly  desirable  that  this
important  concept  be  formulated  purely  on   purely   lattice
language. Some hypothetical application of lattice  theories  can
be connected with the future non-perturbative string theory.  It
is a common folklore [8,9] that  the  total  String  Phase  Space
contains conformal theories as fixed points, integrable models as
interpolating trajectories  between  them,  and  all  the  others
(lattice, q-deformed, etc...) filling in the remnant holes.  In
connection with this, an  interesting  question  of  relationship
between the various types of q-deformations  and  'latticization'
naturally arised last time [4,10-12].

     Up to now, only the lattice analogue of Virasoro algebra was
studied [1-3]. The aim of present paper  is  to  demonstrate  the
general method of obtaining lattice $W$-algebras ($LW$) analogous  to
classical series $WA$, $WB$, $WD$. Further on lattice $W{\cal G}$
  algebra  will
be denoted as $LW{\cal G}$. The method is effective on both the  classical
and  quantum  levels,  although  the  explicit  calculations  are
presented only for the  classical  limit  of $LW_3\equiv LW{\cal A}_2$.
 We  also discuss the relationship between Faddeev-Takhtadjan-Volkov  (FTV)
lattice Virasoro  algebra  and  Lie  algebraic  q-deformation  of
Virasoro algebra, obtained by the authors recently [12].

     For the reminder we bring here some  basic  definitions  and
formulas concerning the theory of classical series of $W$-algebras.
As our main  interest is in studying  minimal  models,  we
discuss in this paper only those $W$-algebras possessing maximally
degenerated representations [13]. So we
consider  some  semisimple  algebra  ${\cal G}$  and  associated  with  it
$W{\cal G}$-algebra generated  by  the  set  of  currents $\{A^p(z)\}$,
where
$p$ spin.

Denote the system of simple roots as $\Phi_s={\{\alpha_r\}}_{r=1}^l$
and consider the theories with energy-momentum tensors (EMT) of the form:

$$T(z)=-t_{ij}(\partial \phi_i \partial \phi_j)(z)+2i\alpha_0 \rho \cdot
\partial^2 \phi(z) \eqno(1)$$
where $t_{ij}$ is a restriction  of  Killing  form  to  the  Cartan
subalgebra and $\rho$ is a halfsum of positive roots. Round brackets
as usual mean normal ordering.
 Minimal  model ${\cal M}_{p,q}$ is described by the EMT with  central
charge
$$c_{p,q}=l\left(1-{\tilde h (\tilde h +1)(p-q)^2\over pq}\right)$$

The main property of these models is  that  all  the  correlation
functions of the primary fields can be calculated in the  Coulomb
gas representation after introducing of the  so-called  screening
charges (SC). They are defined as follows:

$$Q_r^{\pm}=\oint{dz\over 2\pi i}(e^{i\alpha_{\pm}\alpha_r\cdot\phi})(z)$$
where $\alpha_{\pm}$ is determined from the condition that the conformal
dimension of the integrand be equal to one:
$$\alpha_{\pm}^2-2\alpha_0\alpha_{\pm}-1=0$$
The crucial facts about SC's we use are [13-14]: (i) they  form  the  root
part  of  Cartan-Weyl  basis   of quantum  algebra  ${\cal U}_q({\cal G})$
with $q=exp(i\pi \alpha_+^2)$ ; (ii) they commute with all the generators
of
$W{\cal G}$-algebra.
In order to define the right lattice analogue of  $W{\cal G}$-algebra,  we
should consider the lattice versions of  these  statements.  This
will be done in next section.

\section{ General prescriptions}

     In this section we give the basic formulas  and  definitions
concerning the lattice minimal  models.  The  naturality  of  the
method first proposed by Feigin [15] is guaranteed  by  its  full
parallelism with  continuous  case.  Consider  for  example,  the
relation between the two screened vertex operators (VO) corresponding
to simple roots:
$$V_{\alpha_i}^{\pm}(x)V_{\alpha_j}^{\pm}(z)
=(x-z)^{\alpha_{\pm}^2 A_{ij}}(V_{\alpha_i}^{\pm}(x)
V_{\alpha_j}^{\pm}(z))\eqno(2)$$

where $A_{ij}$ is Cartan matrix.
In order to obtain its lattice analog introduce first the natural order in
$\Phi_s$. For the case of ${\cal A}_l$ we have $\alpha_i=e_i-e_{i+1}$.
Then we put lattice VO on $l$ slightly shifted adjacent lattices
such that $a_i(n)$ be left with respect to $a_j(n)$ if $i < j$. Thus
lattice version of the eq.(2) is

$$a_i(m)a_j(n)=q^{A_{ij}}a_j(n)a_i(m),\hbox{\ when\ }n>m\eqno(2^{\prime})$$
\centerline{and}
$$a_i(n)a_{i+1}(n)=q^{-1}a_{i+1}(n)a_i(n)$$
$$a_i(n)a_j(n)=a_j(n)a_i(n),\hbox{\ when\ }\mid i-j\mid\geq 2$$

Note that $q$ now can be treated as free  parameter  contrary  to the
continuous case where it was rigidly determined by the formula in
the end of Introduction.

     Now we introduce a lattice analogue of the SC as

$$Q_i^{\pm}=\sum_n(a_i(n))^{\pm 1}\eqno(3)$$

and define $LW{\cal G}$-algebra as {\bf zero-graded part of  kernel of the
adjoint
action of lattice SO's}, where gradation is defined via the rules
($\bar a\equiv a^{-1}$):
        $$ deg(a_r(n))=1 \;\;\;     deg({\bar a}_r(n))=-1$$

The most common anzats satisfying the condition stated  is  given
by the following series $(LW_3 -case)$

$${\cal L}=\sum_{k,p,r,s}f_{k,p,r,s}{{\bar b}_n}^k{\bar c}_n^{\,p}
b_{n+1}^{k+r}
c_{n+1}^{p+s} {\bar b}_{n+2}^r {\bar c}_{n+2}^{\,s}\eqno(4)$$
\bigskip
$${\cal W}=\sum_{k,p,r,s,t,u}g_{k,p,r,s,t,u}{{\bar
b}_n}^k c_n^p b_{n+1}^{k-r}
{\bar c}_{n+1}^{\,p-s} {\bar b}_{n+2}^{t-r}
 c_{n+2}^{u-s} b_{n+3}^t {\bar c}_{n+3}^{\,u}\eqno(5)$$

where basic commutation relations (2') here take the form
$$b_n\, b_{n+\Delta }=q^2b_{n+\Delta }\, b_n\;\;\; c_n\,c_{n+\Delta}=
q^2c_{n+\Delta }\,c_n$$

$$b_n\, c_{n+\Delta }=q^{-1}c_{n+\Delta }\, b_n\;\;\;
b_n\,c_n=q^{-1}c_nb_n$$

Commuting these two operators with the  SC $Q_b$ one obtains the
following Quantum Master Equations ({\bf QME})

$$-q^{-(k+1)}[k+1]f_{k+1,p,r,s}+q^{k+r}f_{k,p,r,s}-q^{-(r+1)}
[s-r-1]f_{k,p,r+1,s}=0$$

$$q^{-(k+1)}[k+1]g_{k+1,p,r,s,t,u}+q^{k-r}[k+p+r]g_{k,p,r,s,t,u}+$$
$$+q^{r-t-1}[s+r+t-1]g_{k,p,r-1,s,t,u}+q^{t-1}
[t+u-1]g_{k,p,r-1,s,t-1,u}=0\eqno(6)$$
Commuting with $Q_c$ gives two conjugated equations differring from the
above by the replacement $(k,r)\leftrightarrow (s,p)\hbox{\ and\ }
(k,r,t)\leftrightarrow (s,p,u)$.

QME for lattice  generators $A^p(n)$ forming higher $LW$-algebras can be
obtained from the quite analogous considerations. We do not bring here the
corresponding cumbersome expressions for generators and their QME's,
saying only that the single rule being used to obtain them is that
 each monomial entering the formula like (4) for spin-$p$ generator should
 be a zero-graded $p$-local expression, i.e. including the operators from
 $a_i(n)$ to $a_i(n+p)$ for all $i$.  We give here the solutions of QME's
(6) for the reader could imagine what the matter is. Functions $f$ and $g$
as well as their higher analogues differ from zero in some half-infinite
region, however it turns out that QME cannot be solved through the
characteristiscs method, as it gives only zero-part of the solution. The
intuition for solving QME is provided by their classical limit, which will
be discussed in details in Sect.3. In the formula below $\qbinom xy$ stands
for quantum binomial coefficient ${[x]!\over [y]!\,[x-y]!}$.
$$f_{k,p,r,s}=\left(1+{[k+p]\,[r]\over
[p]\,[s]}q^{k-r-s}\right)(-1)^{k+p+r+s}
q^{(k-r)^2+(s-p)(s-p+1)-pr}$$
$$\cdot \qbinom {k+p-1}{k}\cdot \qbinom {r+s-1}{r}\,\eqno(7)$$
when $k,\ p,\ r,\ s\ >0$ and analogous rather cumbersome expression for
 $g_{k,p,r,s,t,u}$. Formally looking at eq.(7), one sees that it is not
invariant under the replacement of the indices mentioned above. In fact,
this can be removed by the appropriate redefinition of $f$ on the boundary
of the region $k,\ p,\ r,\ s\ >0$.

\section{ $LW_3$-algebra in the classical limit}

We begin this section from defining the classical limit of the basic VO's
$b_n$ and $c_n$ and introduce more convenient variables
$$p_n \equiv {\bar b}_n b_{n+1}\;\;\; \hbox{and}\;\;\; d_n \equiv
c_n{\bar c}_{n+1}$$
Basic commutators in the classical limit become
$$\{b_n,b_m\}=\{c_n,c_m\}=\epsilon (n-m)$$
$$\{b_n,c_m\}=\theta (n-m)$$
where $\epsilon (n)$ and $\theta (n)$ are standardly defined as
$$\epsilon (n)=\cases{1, &if $n>0$\cr 0, &if $n=0$\cr -1, &if $n<0$}\qquad
 \theta (n)=\cases{1, &if $n\ge 0$\cr 0, &otherwise}$$
Searching for the generators of $LW_3$ in the form (5) from the condition
 of their commutativity with SC one obtains Classical Master Equations by
 simply omitting factors of the type $q^{something}$ and replacing quantum
numbers
by usual ones in QME's (6). Solving these simplified equations, one readily
obtains the classical versions of the series like (7). Because of absence
of
unpleasant $q$-factors these series can be converted into finite
expressions.
This allows one to calculate the algebra explicitly. Operators $L_n$ and
$W_n$ of
"spin" (=locality) 2 and 3 respectively are given by the following
expressions:
$$L_n=(p_{n+1}d_n+p_{n+1}+d_n)(1+p_n+d_n)^{-1}(1+p_{n+1}+d_{n+1})^{-1}\eqno(
8)$$
$$W_n=d_np_{n+2}(1+p_n+d_n)^{-1}(1+p_{n+1}+d_{n+1})^{-1}(1+p_{n+2}+d_{n+2})^
{-1}
\eqno(9)$$

After a straightforward but rather tedious calculation one can find that
these
operators form the following closed algebra:
\eject
$$\{L_n,L_{n+1}\}=(L_nL_{n+1}-W_n)(1-L_n-L_{n+1})\;\;\;
\{L_n,L_{n+2}\}=-L_nL_{n+1}L_{n+2}+W_nL_{n+2}+W_{n+1}L_n$$
$$\{L_n,W_n\}=\{W_n,L_{n+1}\}=-W_nL_nL_{n+1}+W_n^2$$
$$\{W_n,L_{n+3}\}=-W_nL_{n+2}L_{n+3}+W_nW_{n+2}\;\;\;
\{W_n,L_{n+2}\}=W_nL_{n+2}(1-L_{n+1}-L_{n+2})+W_nW_{n+1}$$
$$\{L_n,W_{n+1}\}=L_nW_{n+1}(1-L_n-L_{n+1})+W_nW_{n+1}\;\;\;
\{L_n,W_{n+2}\}=-L_nL_{n+1}W_{n+2}+W_nW_{n+2}$$
$$\{W_n,W_{n+1}\}=W_nW_{n+1}(1-L_n-L_{n+2})\;\;\;
\{W_n,W_{n+2}\}=W_nW_{n+2}(1-L_{n+1}-L_{n+2})$$
$$\{W_n,W_{n+3}\}=-W_nW_{n+3}L_{n+2}\eqno(10)$$

Strange as it might seemed that operators $L_n$ do not form a subalgebra in
$LW_3$.
However, looking attentively at the first line of commutation relations
(10), one can
see that setting $W_n=0$ one obtains FTV algebra [1].

\section{ Lattice Boussinesq Hierarchy}

	In this section we consider the direct lattice analogue of Boussinesq
	hierarchy, related to classical $W_3$ algebra in the continuous limit.
	First we define the notion of lattice hamiltonian. Fundamental
requirement for an operator to be called hamiltonian is that it commute
with some auxiliary $sl(3)$-algebra. However, in this paper we will
elaborate with more indirect, but convenient criteria: hamiltomians which
will be defined below are in involution with each other and give the set of
equations of motion (EM) becoming the well-known Boussinesq hierarchy in
the
continuous limit.

	Define the zeroth hamiltonian as
	$${\cal H} \equiv {\cal H}^{(0)} ={1 \over 3}\sum_n lnW_n\eqno(11)$$
The pair (${\cal H}$,{\it Poisson Bracket} (10)) generates the following
EM:
$${\dot{L}}_n=\{{\cal H}, L_n\}=(W_{n-1}-W_n)+L_n(L_{n+1}-L_{n-1})$$
$${\dot{W}}_n= \{{\cal H}, W_n\}=W_n(L_{n+2}-L_{n-1})\eqno(12)$$

It can be shown, that the system (12) is completely integrable, however
in this paper we explicitly demonstrate some weaker property that it is
bi-hamiltonian and there exists a recursion operator, allowing one to
built the infinite set of integrals of motion (IM). Below the several
first IM are presented:
$${\cal H}^{(1)}=\sum_n L_n$$
$${\cal H}^{(2)}=\sum_n({L_n^2 \over 2}+L_nL_{n+1}-W_n-L_n)\eqno(13)$$
The interesting points, related to correct
continuous limit will be discussed in more details in next section.

\section{ Bi-hamiltonian structure of Lattice Boussinesq Hierarchy (LBH)}

In this section we demonstrate the existence of two  pairs {\it
(Hamiltonian, Symplectic Structure)},
consistent with each other in the following sense:
$$\dot{\Psi}={\{{\cal H}^{(0)},\Psi \}}_2={\{{\cal H}^{(1)},\Psi \}}_1$$
where $\Psi$ is any operator from the set $\{L_n,W_n \}$. Then the
recursion operator ${\cal R}:\,\,{\cal H}^{(n)}\,\longrightarrow {\cal
H}^{(n+1)}$
has the form
$${\cal R}=\Omega_1^{-1} \Omega_2$$
where $\Omega_i$ is formal denotion of $i$-th symplectic structure.
We recall the standard rules of building the first structure.
Consider the Miura map ${\cal M}:\,\,\{\lambda_n, \omega_n\}\ni
\psi \,\longrightarrow \Psi \in \{L_n,W_n\}$, where $\{\lambda_n,
\omega_n\}$
 is some set of variables with given Poisson bracket $\Omega$. Clearly,
  $\Omega_2$ is uniquely determined by $\Omega$ via the Miura map as
  $$\Omega_2[\Psi_1,\Psi_2]:=\Omega [{\cal M}(\psi_1),{\cal M}(\psi_2)]$$
  Quite analogous formula can be written for the definition of the first
   structure:
   $$\Omega_1[\Psi_1,\Psi_2]:={\cal M}(\Omega [{\cal M}^{-1}(\Psi_1),
   {\cal M}^{-1}(\Psi_2)])\eqno(14)$$

   It can be shown by direct calculation that the following definition of
the basic structure $\Omega$
$$\{\lambda_n,\lambda_{n+1}\}=\lambda_n\lambda_{n+1}-\omega_n\,\,\,\,
\{\omega_n,\omega_{n+1}\}=\omega_n\omega_{n+1}$$
$$\{\lambda_{n+2},\omega_n\}=-\lambda_{n+2}\omega_n\,\,\,\,
\{\lambda_n,\omega_{n+1}\}=\lambda_n\omega_{n+1}$$
$$\{\omega_n,\omega_{n+2}\}=\omega_n\omega_{n+2}\eqno(15)$$
with the explicit realization
$$\lambda_n=p_n+d_n\;\;\; \hbox{and}\;\;\; \omega_n=p_nd_{n+1}$$
provides the bi-hamiltonian structure of the hierarchy (12-13).

Taking ${\cal H}^{(0)}$ and ${\cal H}^{(1)}$ from the previous section,
$\Omega_2$ from eq.(10) and $\Omega_1$ from (14) one obtains the
bi-hamiltonian form of LBH.

Here we would like to drop reader's attention to the following interesting
property
of the first structure $\Omega_1$ related to the explicit form of
$LW_3$-algebra (10).
 Namely, $\Omega_1$ can be viewed as a  {\it contraction} of $\Omega_2$ if
we introduce the following natural gradation (deg($operator$)=$spin$-1)
$$deg(L_n)=1\;\;\;\; deg(W_n)=2\;\;\;\; deg(\Omega_2)=0$$
and rewrite the algebra (10) in terms of graded operators $\epsilon L_n$
and
$\epsilon^2W_n$. One can easily check that
$$\hbox{Eqs}.(15)=lim_{\epsilon \rightarrow 0} \hbox{Eqs}.(10)$$
It is noteworthy, that the same property is valid for the bi-hamiltonian
structure of Volterra model. The details on this point will be given
elsewhere [18].
We conclude this section by remark concerning the continuous limit. Note,
firstly,
that the operator $W_n$ as defined by (9) consists in fact
of two parts of locality 3 and 2. This means that the true spin-3-field is
some
appropriate combination of $W_n$ and $L_n$. Secondly, in the continuous
limit the operators
(9) can be expanded in the series on $\Delta$, which begin from the
constant. This means that
the correspondent Hamiltonians should be appropriately regularized before
taking
the limit. Straightforward
calculation gives the following expansions ($x\equiv n\Delta$)
$$W_n\;\longrightarrow \;{1 \over 27}(1-{\Delta}^2u(x)-{{\Delta}^3\over
2}w(x))$$
$$L_n\;\longrightarrow \;{1\over 3}(1-{{\Delta}^2\over 3}u(x))$$
where $u(x)$ and $w(x)$ are the fields forming usual classical $W_3$
algebra.
Thus if we define, for example
 $${\tilde{\cal H}}^{(0)} \equiv {6\over {\Delta}^3}\sum_n(L_n-{1\over
3}lnW_n-
{1\over 3}-ln3)$$
 $${\tilde{\cal H}}^{(1)} \equiv -{9\over {\Delta}^2}\sum_n(L_n-{1\over
3})$$
then  ${\tilde{\cal H}}^{(0,1)}$ become usual Boussinesq integrals
$${\tilde{\cal H}}^{(0)} \longrightarrow  \int dx\; w(x)$$
$${\tilde{\cal H}}^{(1)} \longrightarrow  \int dx\; u(x)$$
and ${\tilde{\cal H}}^{(0)}$ in continuous limit generates well-known
Boussinesq equations [17]
$$\dot{u}=-u_{xx}+2w_x\;\;,\quad \dot{w}=w_{xx}-{2\over 3}u_{xxx}-{2\over
3}uu_x$$

\section{ Q-deformation as "latticization"}

In this section we are going to discuss the question, lying partly aside,
but immediately related to the problem under consideration. In their recent
paper on Lie-algebraic $q$-deformations of Virasoro algebra [12] authors
demonstrated that this operation, leading to splitting of multiple poles
in quantum operator product expansion (OPE), considered in classical limit
means
authomatic transition from continuum to the lattice with step $\Delta$, if
the deformation parameter is given by $q=exp(2\pi i\Delta )$. Below we
give a short list of interesting formulas and try to understand the
relation between the algebras discussed in previous sections and in [12].

Lie-algebraic $q$-deformations of Virasoro and Superconformal algebras
were found by Chaichian and Presnajder [16] and independently by the
present authors [12]. These algebras naturally originate from the free
fields models. We discuss here only the q-deformation of $Vir$. Consider
free bosonic field $a_p$ with
$q$-deformed commutation relations
$$[a_p,a_s]=[p]\delta_{p+s,0}$$

The following family of operators can be built from $a_n$ :
$$l_p^{\alpha}=\sum_s \{\alpha (2s+p)\}\, a_{-s}a_{p+s}\eqno(16)$$
where curl brackets denote  the  natural  "fermionic"  analog  of
usual quantum number $\{x\}\equiv {q^x+q^{-x}\over q+q^{-1}}$.
Direct calculation shows that $l_p^{\alpha}$ form two-loop Lie algebra
$$[l_p^{\alpha},l_s^{\beta}]=[(\beta +1)p-(\alpha +1)s]l_{p+s}^{\alpha
+\beta +1}-
[(\beta -1)p-(\alpha -1)s]l_{p+s}^{\alpha +\beta -1}+$$
$$+[(\beta +1)p+(\alpha -1)s]l_{p+s}^{\alpha -\beta -1}-
[(\beta -1)p+(\alpha +1)s]l_{p+s}^{\alpha -\beta +1}+\hbox{\ central terms\
}\eqno(17)$$
We have omitted central extension as it is irrelevant in the given context.
We note
 only that in the limit $q \rightarrow 1$ the central term tends to right
expression
 $p(p^2-1)$ (see [12] for detailed discussion on this point)
Quantum OPE has the form ($T^{\alpha}(x)\equiv \sum_p
l_p^{\alpha}x^{-p-2}$):
$$T^{\alpha}(z)\,T^{\beta}(w)={T^{\alpha +\beta +1}(wq^{\alpha +1})\over
z-wq^
{\alpha +\beta +2}}-{T^{\alpha +\beta +1}(wq^{-\alpha -1})\over
z-wq^{-\alpha
-\beta -2}}+\ldots $$
When going to the classical limit it becomes
$${\{T^{\alpha}(x),T^{\beta}(y)\}}_{P.B.}=
T^{\alpha +\beta +1}(y+\Delta (\alpha +1))\delta (x-y-\Delta (\alpha +\beta
+2))-$$
$$-T^{\alpha +\beta +1}(y-\Delta (\alpha +1))\delta (x-y+\Delta (\alpha
+\beta +2))+\ldots$$
Thus one observes that  new parameter $\Delta $ naturally appears in the
model as
a shift on coordinate space and should obviously be interpreted as a
lattice constant.
Indeed, it is convenient to introduce a lattice with step $\Delta$ and to
rewrite
the previous equation as follows
$$\{ \Lambda_n^{\alpha},\Lambda_m^{\beta} \}=\Lambda_{m+\alpha +1}^{\alpha
+
\beta +1}\,\delta_{n,\, m+\alpha +\beta +2}-\Lambda_{m-\alpha -1}^{\alpha +
\beta +1}\, \delta_{n,\, m-\alpha -\beta -2}+\ldots \eqno(18)$$
where
$$\Lambda_n^{\alpha}\;\equiv \int_{n\Delta}^{(n+1)\Delta} dx\;
T^{\alpha}(x)$$

Superconstuction bases on the fact that the same algebra (17) is formed by
the following bilinear operators
$$M_p^{\alpha}=\sum_s [\alpha (2s+p)]\, \psi_{-s}\psi_{p+s}\eqno(19)$$
where $\psi_p$ is a fermion field with deformed anticommutator
$$\{\psi_p,\psi_s\}=\{p\}\delta_{p+s,0}$$

However, if one considers the same object (19) built from usual fermions
then one obtains another two-loop algebra, {\it non-isomorphic} to (17).
In somewhat analogous property takes place in the zoo of lattice algebras.
 Analogues of Vir-operators entering the $LW$-algebras even do not form
 closed subalgebras. However, as it follows from the remark in the end of
Sect.3,
one can hope, that $LW_k$-algebras with $k<n$ can be {\it extracted} from
$LW_n$-algebra by setting to zero generators with spins $>k$. In any case,
it seems
that the lattice analogue of the notion of conformal invariance, if
existing, cannot
 be formulated in purely algebraic terms. From the other side, the
remarkable property
of FTV-algebra in classical limit  found in [12] is
 that it can be rewritten as a Lie algebra isomorphic to that of (18).

\end{document}